\documentclass[sigconf,nonacm]{acmart}

\renewcommand\footnotetextcopyrightpermission[1]{}
\setcopyright{none}
\usepackage{booktabs}
\usepackage{tabularx}
\usepackage{array}
\usepackage{enumitem}
\usepackage{microtype}

\AtBeginDocument{%
}

\acmConference[Converted Manuscript]{Converted Manuscript}{2022}{Beijing, China}
\acmYear{2022}
\acmISBN{}
\acmDOI{}

\title{Retrieval of Scientific and Technological Resources for Experts and Scholars}

\author{Suyu Ouyang}
\email{darkham97@163.com}
\affiliation{%
  \institution{School of Computer Science (National Pilot School of Software Engineering), Beijing University of Posts and Telecommunications}
  \institution{Beijing Key Laboratory of Intelligent Telecommunication Software and Multimedia}
  \city{Beijing}
  \country{China}
}

\author{Yingxia Shao}
\authornote{Corresponding author.}
\affiliation{%
  \institution{School of Computer Science (National Pilot School of Software Engineering), Beijing University of Posts and Telecommunications}
  \institution{Beijing Key Laboratory of Intelligent Telecommunication Software and Multimedia}
  \city{Beijing}
  \country{China}
}

\author{Ang Li}
\affiliation{%
  \institution{School of Computer Science (National Pilot School of Software Engineering), Beijing University of Posts and Telecommunications}
  \institution{Beijing Key Laboratory of Intelligent Telecommunication Software and Multimedia}
  \city{Beijing}
  \country{China}
}

\begin{document}

\begin{abstract}
Institutions of higher learning, research institutes and other scientific research units have abundant scientific and technological resources of experts and scholars, and these talents with great scientific and technological innovation ability are an important force to promote industrial upgrading. The scientific and technological resources of experts and scholars are mainly composed of basic attributes and scientific research achievements. The basic attributes include information such as research interests, institutions, and educational work experience. However, due to information asymmetry and other reasons, the scientific and technological resources of experts and scholars cannot be connected with the society in a timely manner, and social needs cannot be accurately matched with experts and scholars. Therefore, it is very necessary to build an expert and scholar information database and provide relevant expert and scholar retrieval services. This paper sorts out the related research work in this field from four aspects: text relation extraction, text knowledge representation learning, text vector retrieval and visualization system.
\end{abstract}


\maketitle

\section{Introduction}

In recent years, with the vigorous introduction of policies related to technological innovation-driven development, technological innovation and management innovation have increasingly become an important way of social development and economic benefits. The private economy is an important part of the national economy, and small and medium-sized enterprises are the main part of my country's private economy. Therefore, how to effectively meet the needs of society and improve the innovation ability of private enterprises, especially small and medium-sized enterprises, has become an important research topic in my country's economic development. Related studies on innovation and entrepreneurship also show that creative personality, education and entrepreneurial identity jointly influence innovation behavior \cite{zhou2020creative}. Small and medium-sized enterprises are naturally at a disadvantage in terms of talent competition and technological innovation due to their relatively weak capital reserves and lack of innovation teams. In the process of product research and development, when faced with technical difficulties, enterprises are limited by the above reasons and often seek relevant resources through their own social relations. This method is inefficient and expensive, and the experts found may not be able to really meet the needs of the enterprise.

Institutions of higher learning, research institutes and other scientific research units have abundant scientific and technological resources of experts and scholars, and these talents with great scientific and technological innovation ability are an important force to promote industrial upgrading. The scientific and technological resources of experts and scholars are mainly composed of basic attributes and scientific research achievements. The basic attributes include information such as research interests, institutions, and educational work experience. However, due to information asymmetry and other reasons, the scientific and technological resources of experts and scholars cannot be connected with the society in a timely manner, and social needs cannot be accurately matched with experts and scholars. Therefore, it is very necessary to build an expert and scholar information database and provide relevant expert and scholar retrieval services.

Because the scientific and technological resources of experts and scholars are subdivided and difficult to cover completely, retrieval of these resources needs to mine the relationship between text semantics, integrate them to form a knowledge map of professional fields, and present them efficiently in a visual way. Scholar-resource modeling has therefore begun to consider multi-view scholar clustering and dynamic interest tracking, so as to capture evolving research interests from heterogeneous evidence \cite{li2023multiViewScholar}. In addition, different from the traditional keyword matching method, the retrieval of scientific and technological resources of experts and scholars needs to meet the characteristics of approximate retrieval. How to carry out feature semantic learning on the basis of existing scientific and technological resources, establish a common semantic representation space of scientific and technological resources from different sources, and realize accurate and approximate search of scientific and technological resources retrieval by experts and scholars based on the ranking and scoring mechanism is of great practical significance. At the infrastructure level, scalable random-walk frameworks over billion-edge natural graphs provide useful support for large-scale academic graph mining \cite{shao2021memoryAware}.

\section{Text Entity Relation Extraction}

Information extraction is to extract structured information from text, and text entity relation extraction is a sub-domain of information extraction, which refers to the extraction of semantic relations from text relations, and this semantic relation exists between entity pairs. Defining text $S$, relation set $R = \{r_1, r_2, \ldots\}$, text relation extraction is the process of extracting triples based on $(h, r, t) \in R$, where $h$ represents the head entity, $t$ represents the tail entity, and $r$ represents the relationship between $h$ and $t$ \cite{liu2016knowledge}. Generally speaking, the steps of text relation extraction are divided into named entity recognition (Named Entity Recognition, NER) \cite{nadeau2007survey} and relation classification (Relation Extraction, RE) \cite{rink2010utd}. In deep learning, text relation extraction is mainly studied from two methods based on supervised and remote supervision. Supervised relation extraction methods mainly include pipeline learning and joint learning.

Pipeline extraction usually separates named entity recognition and semantic relationship classification. It first identifies the entities existing in the text data, and then judges whether the relationship exists according to the entity pair. Early pipeline learning methods were mainly based on convolutional neural networks (Convolutional Neural Networks) \cite{li2017recursive}. Zeng et al. \cite{zeng2014relation} used a CNN network for relation classification. The design was based on lexical-level and sentence-level feature extraction networks; the two features are fused to obtain the encoding vector, the fully connected layer is followed by the encoder network, and softmax is used to classify the relationship. The effectiveness of the method is verified on the public data set. On this basis, Nguyen et al. \cite{nguyen2015relation} added multi-scale convolution kernels, fully using sentence-level features, and can automatically learn the implicit features in sentences. In recent years, many scholars have carried out research based on Recurrent Neural Network (RNN). Socher et al. \cite{socher2012semantic} used RNN to syntactically parse text for the first time. Each node on the syntactic parse tree consists of a vector and a transformation matrix. For words and sentences of any syntactic type and length, the combined vector representation can be learned. Long Short-Term Memory Network (LSTM) is a special RNN. Xu et al. \cite{xu2015classifying} proposed the SDP-LSTM model for relation classification, based on the idea of the shortest dependency path, filtering the useless information of text, using LSTM to effectively integrate heterogeneous information, and formulating an effective dropout strategy to prevent the possibility of overfitting. Zhang et al. \cite{zhang2015bidirectional} proposed BRCNN, based on the shortest dependency path, using a bidirectional long short-term memory network (Bi-LSTM), considering the directionality of the relationship between entities, and using the information before and after the word to extract the relationship.

However, pipelined methods treat named entity recognition and relationship classification as two separate tasks, ignoring the relationships that exist between tasks. At the same time, due to the sequence between tasks, the error of entity recognition will affect the subsequent relationship classification, and the phenomenon of error propagation will occur. In addition, since entity pairs lacking relationships cannot be paired in pairs, redundant information will be brought into the relationship classification, and entity pair redundancy will appear. In recent years, many studies have attempted to jointly learn the two tasks, that is, fusing named entity recognition and relation classification into a single task. Joint learning methods are mainly divided into methods based on parameter sharing, methods based on sequence annotation and methods based on graph structure.

\begin{enumerate}[label=(\arabic*),leftmargin=*]
\item The method based on parameter sharing is to design a shared coding layer between the two tasks of named entity recognition and relation classification, and obtain the best global parameters during training. Miwa et al. \cite{miwa2016end} first designed LSTM-based shared encoding layers to obtain substructure information on word sequences and syntactic dependency trees. On this basis, Zheng et al. \cite{zheng2017hybrid} designed a shared coding layer based on the vector embedding layer and the Bi-LSTM layer, the named entity recognition module used LSTM, and the relation classification module used CNN, which effectively captured the long text entity labels and distance dependence.

\item The method based on sequence labeling is the problem of combining the two tasks of named entity recognition and relation classification into sequence labeling. In order to solve the problem that the method based on parameter sharing is prone to generate entity pair redundancy, Zheng et al. \cite{zheng2017tagging} proposed a new labeling scheme, which transformed the joint extraction task into a labeling problem, and studied the relationship extraction performance between different end-to-end models. A biased loss function was additionally used to enhance associations between related entities. Bekoulis et al. \cite{bekoulis2018joint} used conditional random fields (Conditional Random Field, CRF) to model the task of named entity recognition and relation extraction as a multi-head selection problem (Multi-Head Selection Problem), regarding the relationship classification task as multiple binary classification tasks, so that each entity can judge the relationship with all other entities, and effectively solve the problem of relationship overlap.

\item The method based on graph structure \cite{shi2019deep} is to use graph neural network (Graph Neural Networks, GNN) to extract relations. Wang et al. \cite{wang2018joint} proposed a joint learning model based on a graph structure, while using a loss function with biased weights to weaken the influence of invalid labels and enhance the association between related entities. In addition, Fu et al. \cite{fu2019graphrel} proposed an end-to-end relation extraction model GraphRel, through relational weighted graph convolutional networks (Graph Convolutional Network, GCN) to consider the interaction between entities and relations, combining RNN and GNN to extract sequential features and position-dependent features of each word, effectively solving the problem of overlapping entity pairs. Heterogeneous graph attention has also been used to capture semantic and structural signals in short text classification \cite{hu2019heterogeneousGraphAttention}, teacher-student graph distillation can address incomplete graph features and structures \cite{huo2023t2gnn}, and reciprocal contrastive learning has been explored for heterogeneous graph neural networks \cite{jin2022heterogeneousGraph}.
\end{enumerate}

Methods based on remote supervision can greatly reduce labor costs, consume less time, and have strong field portability. Mintz et al. \cite{mintz2009distant} proposed DS-logistic, which uses an external knowledge base to associate a large-scale knowledge graph with text, extracts textual semantic features from remotely supervised annotated data, and inputs its representation into a classifier for relation classification. Entity and relationship understanding has also been extended to downstream tasks such as sarcasm detection and fake news detection \cite{wang2025sarcasm,dong2023deepMarkov}. However, due to the use of automatic labeling, the possibility of errors in the training set data is greatly increased, resulting in a low accuracy rate of the current remote supervision entity relationship extraction.

\section{Text Knowledge Representation Learning}

In recent years, large-scale knowledge graphs such as WordNet and Freebase have provided an effective foundation for important areas of artificial intelligence such as semantic search, intelligent question answering, and personalized recommendation. Knowledge graphs are usually represented using triples $(h, r, t)$, where $h$ and $t$ represent the head entity and tail entity, and $r$ represents the relationship. Knowledge graphs are often constructed manually through databases, and often underperform when learning new knowledge. Knowledge representation learning, also known as knowledge graph embedding, realizes semantic feature extraction of entities and relationships by embedding entity relationships into vector space, and can efficiently calculate entities, relationships and complex semantic associations between them. For distributed information networks, federated self-adaptive learning offers a privacy-aware path for representation learning across separated data owners \cite{li2026fedsin}. In recent years, translation models are a representative method in knowledge representation learning.

Inspired by Word2Vec \cite{mikolov2013distributed}, Bordes et al. \cite{bordes2013translating} proposed TransE, which embeds entities and relations into a low-dimensional vector space, and treats relations as the translation vector between entity pairs for each triple in the knowledge graph. A distance-based loss function is designed, which effectively extracts the semantic representation of knowledge. Wang et al. \cite{wang2014transh} proposed TransH. For each relation in the knowledge base, it defines the corresponding hyperplane, and projects the head and tail entity vectors onto the hyperplane. The distance formula is consistent with TransE, so that the semantic representations of entities in different relationships are different, and the semantic representations of different entities in the same relationship are similar, which effectively solves the shortcomings of TransE in the face of non-one-to-one relationships. In order to better distinguish the difference between entities and relations, Lin et al. \cite{lin2015learning} proposed TransR, which firstly embeds entity and relation distributions in different spaces, then maps entities into relational spaces, and constructs a distance formula between the two projected entities based on the translation model. On this basis, Ji et al. \cite{ji2015dynamic} considered the diversity of entities and relations, and proposed TransD, which defined two vector spaces: the former representing the semantics of entities or relations, and the latter representing how to map entities from entity space to relational space. Community-oriented graph representation is another useful perspective, and modularity-based deep learning has been explored for community detection \cite{yang2016modularityCommunity}.

In addition, Ji et al. \cite{ji2016transparse} proposed TranSparse, which introduced dynamic sparse matrices. For heterogeneity, the sparsity factor depends on the number of entity pairs corresponding to the relation links; for imbalance, different relation projection matrices are used for head and tail entities. Xiao et al. \cite{xiao2015transa} proposed TransA, which no longer simply considers the distance between vectors, but adopts an adaptive metric score function to add weights to different dimensions of the vector. Furthermore, KG2E \cite{he2015kg2e} and TransG \cite{xiao2015transg} use Gaussian distribution to characterize different semantic information of entities and relations.

In large-scale Internet text data, there are actually more sentences containing entities, and the description information of these entities can help us perform entity representation learning. Reconstructing entity representation using entity description information can effectively improve the relationship prediction task. Socher et al. \cite{socher2013reasoning} proposed NTN, introducing an external news corpus to initialize the representation of each entity by averaging the word vectors contained in its name. Xie et al. \cite{xie2016representation} proposed DKRL, which uses factual text extracted from Freebase as entity description, based on continuous bag of words (CBOW) and deep convolutional neural network models \cite{xu2013image,li2017variance} to obtain the semantics of entity description, and fuses the semantics of entity and text description to learn more about knowledge representation. Wang et al. \cite{wang2016text} proposed TEKE, which utilizes relation mentions and entity descriptions to deal with the ambiguity of relations and entities, and introduces a mutual attention mechanism to mutually enhance the textual representations of relations and entities. Xiao et al. \cite{xiao2017ssp} proposed a semantic space projection model SSP, which embeds triples into semantic subspaces to enhance semantic interaction, and uses text descriptions to discover semantic correlations, which effectively improves the accuracy of semantic embeddings. Scientific publication representation learning further benefits from semantic-similarity attention and hypergraph convolution, which can jointly model publication semantics and higher-order relations \cite{li2026semanticHypergraph}.

With the development of pre-trained language models, work on knowledge representation learning with pre-trained language models such as BERT \cite{devlin2018bert} has gradually increased. Yao et al. \cite{yao2019kgbert} proposed KG-BERT, by modifying the BERT model input to make it suitable for use in the form of knowledge base triples, and achieved great results in tasks such as triple classification, link prediction, and relationship prediction. Liu et al. \cite{liu2020kbert} proposed K-BERT, which effectively solves related tasks in specialized domains by incorporating structured information from knowledge bases into pre-trained models. Zhong et al. \cite{zhong2019knowledge} proposed a knowledge-augmented model KET to understand context through a multi-layered self-attention mechanism, using external knowledge for the problem of emotion recognition in dialogue. In cross-graph scenarios, federated graph neural networks can learn transferable node representations without directly merging graph data \cite{guan2021federatedGraph}.

\section{Text Semantic Vector Representation}

Text semantic vector representation is to use fixed-length vectors to represent sentences of indeterminate length to provide services for downstream tasks such as semantic search \cite{kou2016social}, text clustering \cite{sun2009improvement}, and text classification \cite{minaee2021deep}. Similar representation ideas are also widely used in sequential and session-based recommendation \cite{zhou2022filterEnhanced,xia2021selfSupervised,zhang2025td3}. Text semantic vector representation methods mainly include two methods based on word level and sentence level.

The word level obtains the semantic features of the text by calculating the word vector of the word and weighted average of all the word vectors. Mikolov et al. \cite{mikolov2013distributed} proposed Word2Vec in 2013, using CBOW and Skip-Gram strategies to vectorize all words, effectively mining the relationship between words. On this basis, Pennington et al. \cite{pennington2014glove} proposed GloVe, which combines the global matrix factorization method and the local text box capture method, and is an unsupervised learning algorithm based on word vector representation. Joulin et al. \cite{joulin2016bag} proposed FastText, which adopted the bag-of-words method, introduced the n-gram mechanism, and weighted the average of all words and n-gram vectors to obtain the semantic representation vector of the entire document, and used softmax for multi-classification tasks. Le and Mikolov \cite{le2014distributed} proposed Doc2vec based on Word2vec, which not only considered the semantics between words and words, but also the word order. Common word vector weighting methods include average vector and IDF weighted average. Arora et al. \cite{arora2017simple} proposed SIF, which introduced smooth inverse frequency by computing the weighted average of the word vectors in the sentence and then removing the projection of the average vector on its first principal component. In intelligent decision applications, interpretable machine learning further emphasizes the need for transparent feature representations \cite{li2019interpretableDecision}.

The sentence level regards each sentence as a ``word'', and obtains the sentence vector representation by mining the effective information of the sentence context \cite{hu2018anomaly}. Kiros et al. \cite{kiros2015skip} adopted the idea of Seq2Seq \cite{sutskever2014sequence} and proposed Skip-Thought, which uses the current sentence to predict the previous sentence and next sentence of the current sentence in the article, and generates sentence vectors during model training. On this basis, Logeswaran et al. \cite{logeswaran2018efficient} proposed Quick-Thought, which modified the prediction behavior as a classification problem. Conneau et al. \cite{conneau2017supervised} proposed InferSent, which uses the SNLI dataset to encode sentence pairs using a Bi-LSTM based encoder, using fully connected layers and three-way softmax layers to predict sentence relations. On this basis, Cer et al. \cite{cer2018universal} proposed Universal Sentence Encoder, using Transformer \cite{vaswani2017attention} and DAN \cite{long2015learning} as encoders to extract sentence representations, and introduced a new distance calculation formula for text classification tasks.

BERT \cite{devlin2018bert} has achieved very good results on the task of semantic similarity calculation, but its construction makes it unsuitable for semantic similarity search. Reimers et al. \cite{reimers2019sentencebert} proposed Sentence-BERT, which uses a Siamese neural network architecture: the sentences are input into the BERT model shared by two parameters to obtain the representation vector of each sentence, and finally the representation vector of the sentence is obtained. In order to solve the problem of non-smooth BERT semantics, Li et al. \cite{li2020sentence} proposed BERT-flow, which converts the distribution of sentence vectors into Gaussian distribution, so that the BERT representation embedding is transformed into an isotropic and more uniformly distributed space. Su et al. \cite{su2021whitening} proposed BERT-Whitening, which adopts a simpler linear transformation method. After obtaining the sentence vector, PCA whitening operation is performed on the matrix, so that the mean value of each dimension is 0, and the covariance matrix is a unit matrix. Gao et al. \cite{gao2021simcse} proposed SimCSE, which constructs positive and negative samples in a self-supervised manner to improve the sentence representation ability of the BERT model. Retrieval-oriented pre-training methods such as RetroMAE further adapt masked auto-encoding to dense retrieval scenarios \cite{xiao2022retromae}.

\section{Text Information Retrieval and Visualization}

With the continuous development of deep learning \cite{fang2020identity,lin2009average} in recent years, text information retrieval is gradually divided into two categories: ordinary search and vector retrieval. Ordinary search is based on accurate search content and retrieval sentences. After word segmentation of the document, the disassembled word elements are inverted and indexed through data structures such as KD-Tree, and the text correlation algorithm is used during retrieval. The results are scored and sorted, and the results are finally returned. The data index of vector retrieval \cite{li2022scientific} is different from ordinary search. After the documents, sentences, phrases, etc. are converted into vectors and stored in the search engine, the search engine uses the distance calculation module to cluster and save the vectors. Cross-modal retrieval has also moved toward supervised and federated settings, which is useful when scientific and technological resources include both textual and media information \cite{li2024federatedCrossModal}. Vector retrieval usually occurs in application scenarios where the user is not clear about the specific search term or the user wishes to recall the content pointed to by a wider range of synonyms or synonyms.

Vector retrieval recall is roughly divided into two kinds: linear search and nearest neighbor search (Approximately Nearest Neighbor Search, ANNS). Linear search is an accurate linear brute force search, that is, the vector of the word to be searched is compared with all the vectors in the index in turn, and then sorted by distance. The principle of nearest neighbor search is to reduce the search range of the query vector as much as possible, thereby improving the query speed. It is mainly divided into the following four types.

\begin{enumerate}[label=(\arabic*),leftmargin=*]
\item Tree-based search, which uses a tree structure to represent the division of space. The representative algorithms are KD-Tree and Annoy \cite{spotify2017annoy}. KD-Tree selects the dimension with the largest variance each time it divides the space, and then divides it on this dimension. In contrast, Annoy keeps dividing the vector space with hyperplanes until the number of vectors in each subspace falls below a specified number.

\item Based on the space division of hash, the retrieval vector first performs the hash operation on the query vector to obtain the corresponding bucket, and then searches for similar vectors in the corresponding bucket.

\item Graph-based search. Malkov et al. \cite{malkov2018hnsw} proposed a graph-based data structure HNSW that uses a variant of the greedy search algorithm for approximate nearest neighbor search.

\item Coding based on quantization. The representative algorithm is IVFPQ. IVFPQ divides the high-dimensional space into multiple subspaces, calculates the distance from the word vector to be searched to the cluster center of the subspace, and adds all the distances to obtain an approximate distance.
\end{enumerate}

With the continuous development and application of big data technology, more and more text information retrieval systems \cite{yang2015ontology} have introduced knowledge graph technology to effectively improve user search experience on the basis of traditional search engines. In recent years, large-scale research on knowledge graphs of scientific and technological resources has gradually received more and more attention. The Microsoft Academic Knowledge Graph (Microsoft Academic Graph, MAG) \cite{sinha2015overview} intelligently analyzes academic entities from web pages, and stores the academic knowledge graph constructed by the entities. Tang Jie et al. \cite{tang2008arnetminer} constructed the knowledge graph SciKG based on papers in the computer field, and independently developed the scientific and technological academic literature service platform. AMiner \cite{wang2018acekg} constructed a large-scale academic knowledge graph AceKG in the form of RDF. For incomplete multi-view data, view-category interactive sharing transformers provide another way to exploit complementary views during resource analysis \cite{ou2024viewCategory}.

Elasticsearch is a distributed search and analytics engine based on the Lucene library that has quickly become the most popular search engine since its release in 2010. In recent years, many studies have designed text information retrieval platforms based on Elasticsearch. Jiang Kang \cite{jiang2015metadata} built an index for massive water conservancy heterogeneous data based on Elasticsearch, and designed a water conservancy metadata search and sharing platform. Yu Yue \cite{yu2016adverse} designed a data index and retrieval module based on Elasticsearch, and established an adverse drug reaction knowledge base retrieval system to realize the sharing and retrieval of adverse drug reaction knowledge. Hua Lizhi \cite{hua2018domain} designed a knowledge query and retrieval module based on graph retrieval and Elasticsearch full-text retrieval, built a domain knowledge base management system based on knowledge graph, and solved the problem of information islands. Ye Jifan \cite{ye2020springboot} used SpringBoot to build the project, and used Elasticsearch as the storage engine to design and implement a Chinese social science paper analysis system. Zeng Yafei \cite{zeng2016distributed} designed and implemented a distributed intelligent search engine based on Elasticsearch by combining the vertical domain personalized dictionary construction technology and intelligent recommendation technology. In privacy-sensitive resource systems, federated learning with stochastic quantization can reduce communication while preserving distributed training ability \cite{li2022stochasticQuantization}.

\section{Conclusion}

At present, the retrieval technology of scientific and technological resources for experts and scholars is becoming more and more mature, but researchers still need to invest a lot of energy in continuous exploration. By summarizing the research work on the retrieval of scientific and technological resources of experts and scholars, related research can be carried out from four aspects in future research.

\begin{enumerate}[label=(\arabic*),leftmargin=*]
\item For the task of extracting the relationship between Chinese scientific and technological resources, in Chinese text segmentation, boundary segmentation errors may often occur, resulting in word ambiguity problems. Different word segmentation methods will have different meanings. However, when the existing methods deal with the word segmentation boundary problem, they only use simple jieba word segmentation and other operations, and do not make full use of the position sequence relationship.

\item For the learning task of Chinese scientific and technological text knowledge representation, the existing methods often use random sampling for triple negative sampling, that is, the method of randomly replacing head and tail entities is used for negative sampling, and the generated negative triples often have obvious logical errors. Such low-quality negative samples are less helpful for training.

\item For the task of semantic vector representation of Chinese scientific and technological texts, existing methods often use pre-trained language models in general fields. For texts in professional fields, the performance of language representation is often poor. At the same time, due to the high vector dimension, the efficiency of subsequent computing tasks is low.

\item Continued research and development of scientific and technological resource retrieval and visualization systems for industrial experts and scholars. Through the in-depth integration of academic research and market demand, the reliability, confidence, and execution efficiency of entity relationship extraction are continuously improved, and the performance of the relationship extraction model is further improved, providing more convenience for people's lives.
\end{enumerate}

\begin{acks}
This work is supported by National Key R\&D Program of China (2018YFB1402600), the National Natural Science Foundation of China (61772083, 61877006, 61802028, 62002027).
\end{acks}

\end{document}